%% file: medima-template.tex
\documentclass[times,twocolumn,final]{elsarticle}

\usepackage{medima}
\usepackage{framed,multirow}
\usepackage{soul}
\usepackage{amssymb}
\usepackage{latexsym}
\usepackage{float}
\usepackage{placeins}
\usepackage{makecell}
\usepackage{tabularx}
\usepackage{url}
\usepackage{xcolor}

\usepackage{hyperref}

\definecolor{newcolor}{rgb}{.8,.349,.1}

\journal{Medical Image Analysis}
\input{pre}

\begin{document}

\verso{Feldman \textit{et~al.}}

\begin{frontmatter}

\title{Recursive Variational Autoencoders for 3D Blood Vessel Generative Modeling}%
\author[1,2]{Paula \snm{Feldman}\corref{cor1}}
\cortext[cor1]{Corresponding author: }
\ead{paulafeldman@conicet.gov.ar}
\author[2]{Miguel \snm{Fainstein}}

\author[2]{Viviana \snm{Siless}}
\author[1,3]{Claudio \snm{Delrieux}}
\author[1,2]{Emmanuel \snm{Iarussi}}
\address[1]{Consejo Nacional de Investigaciones Científicas y Técnicas, Argentina}
\address[2]{Universidad Torcuato Di Tella, Buenos Aires, Argentina}
\address[3]{Depto.~Ing.~Eléctrica y Computadoras, Universidad Nacional del Sur, Bahía Blanca, Argentina}

\received{}
\finalform{}
\accepted{}
\availableonline{}
\communicated{}

\begin{abstract}

Anatomical trees play an important role in clinical diagnosis and treatment planning. 
Yet, accurately representing these structures poses significant challenges owing to their intricate and varied topology and geometry.
Most existing methods to synthesize vasculature are rule based, and despite providing some degree of control and variation in the structures produced, they fail to capture the diversity and complexity of actual anatomical data.
We developed a Recursive variational Neural Network (RvNN) that fully exploits the hierarchical organization of the vessel and learns a low-dimensional manifold encoding branch connectivity along with geometry features describing the target surface.
After training, the RvNN latent space can be sampled to generate new vessel geometries. 
By leveraging the power of generative neural networks, we generate 3D models of blood vessels that are both accurate and diverse, which is crucial for medical and surgical training, hemodynamic simulations, and many other purposes.
These results closely resemble real data, achieving high similarity in vessel radii, length, and tortuosity across various datasets, including those with aneurysms.
To the best of our knowledge, this work is the first to utilize this technique for synthesizing blood vessels. 

\end{abstract}


\end{frontmatter}

\footnotetext{Code and live demo are available at~\href{https://github.com/LIA-DiTella/vesselvae}{github.com/LIA-DiTella/vesselvae}.
This is the author's accepted manuscript of a paper accepted for publication in Medical Image Analysis (Elsevier). The final version is available at: [].}

\section{Introduction}

Accurate 3D representations are becoming increasingly essential for modeling complex hierarchical structures such as blood vessels, renal tubules, airways, and similar systems.
These models play an important role in various domains, ranging from disease diagnosis~\cite{roman2012vascular} and prognosis~\cite{murthy2012coronary} to intervention simulation~\cite{le2023rehearsals}. 
Another vital application is surgical planning ~\cite{lawaetz2021simulation}, and fluid dynamics simulation~\cite{taylor2023patient}, among others.
The significance of such detailed models extends beyond the medical domain to fields like computer graphics (i.e. vegetation generation~\cite{lee2023latent, cuntz2010one, banerjee2020sample}). 

Depending on the downstream task, different data structures may be used to effectively represent the 3D model.
Obtaining such representations from patient scans with high resolution and fidelity is not trivial, requires expert medical knowledge, and is highly error-prone~\cite{lan2018re, mou2024costa}.
Despite significant advances in vessel segmentation~\cite{tetteh2020deepvesselnet}, reconstructing thin features accurately from medical images remains challenging~\cite{alblas2021deep}.
As a result, several methods have been developed to adequately synthesize the geometry of blood vessels~\cite{WU20134}.
Moreover, these generative models are often seen as an opportunity to address the scarcity of data on rarer anatomical and pathological variations~\cite{van2024synthetic}.

\begin{figure}[!t]
\includegraphics[width=\columnwidth]{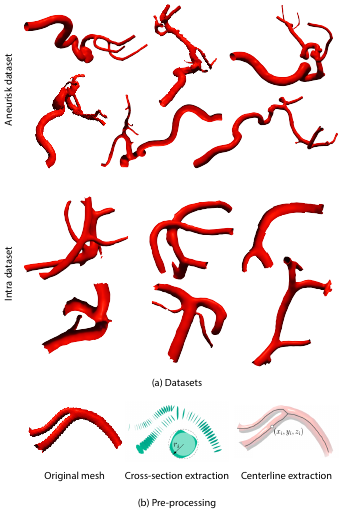}
\caption{Datasets and pre-processing overview: The raw meshes from the Intra 3D collection undergo pre-processing using the VMTK toolkit. This step is crucial for extracting centerlines and cross-sections from the meshes, which are then used to construct their binary tree representations.}
\label{fig:dataset}
\end{figure}

Within the literature on generating vascular 3D models, we identified two main families: fractal-based, and space-filling algorithms.
Fractal-based algorithms use a set of fixed rules that include different branching parameters, such as the ratio of asymmetry in arterial bifurcations and the relationship between the diameter of the vessel and the flow ~\cite{galarreta2013three,zamir2001arterial}. 
On the other hand, space-filling algorithms enable blood vessels to grow within a defined perfusion volume while adhering to hemodynamic laws with constraints related to flow and radius~\cite{hamarneh2010vascusynth,talou2021adaptive,schneider2012tissue,merrem2017computational,rauch2021interactive}. 
Although these \emph{model-based} methods provide some degree of control and variation in the structures produced, they often fail to capture the complexity and diversity of real anatomical data. Additionally, the parameter space exploration for desirable results is very haphazard.

In recent years, deep neural networks led to the development of powerful generative models~\cite{xu2023generative}, such as Generative Adversarial Networks (GANs)~\cite{goodfellow2020generative,kazeminia2020gans}, Diffusion Models~\cite{ho2020denoising}, and Transformers~\cite{vaswani2017attention, siddiqui2024meshgpt}, demonstrating groundbreaking performance in many applications, ranging from image and video synthesis to protein structure synthesis~\cite{ramesh2021zero, yan2021videogpt, jumper2021highly}.
These advances have inspired the creation of novel network architectures to model 3D shapes using voxel representations~\cite{wu2016learning}, point clouds~\cite{yang2019pointflow}, signed distance functions~\cite{park2019deepsdf}, and polygonal meshes~\cite{nash2020polygen}.
However, the use of deep learning to model blood vessel structures has been hindered by the lack of high-quality datasets and the complexity inherent in representing tree-like structured data.

A pioneering approach to generating arterial anatomies with deep generative models was introduced in~\cite{wolterink2018blood}.
However, their model remained confined to single-channel vessels, making it unsuitable for generating more complex topologies. 
More recently, a diffusion model was proposed by~\cite{sinha2024representing}, operating within the space of Implicit Neural Representations to learn vessel tree distributions. 
\cite{deo2024few} also builds on latent diffusion models, particularly few-shot learning for generating aneurysms geometries. 
\cite{kuipers2024generating} 
uses diffusion with a set-transformer architecture to generate vessels as point clouds, but requires a post-processing step to reconstruct the centerlines. \cite{prabhakar20243d} introduced 
 a novel graph-based diffusion model capable of handling cycles, yet it primarily targets capillary networks and is limited to generating the vessels' topology without a corresponding 3D mesh.

In a preceding publication, we presented a framework named VesselVAE~\cite{feldman2023vesselvae} for synthesizing blood vessel geometry.
This generative framework is based on a Recursive variational Neural Network (RvNN), that has been successfully applied in various contexts, including natural language~\cite{socher2011parsing,socher2014recursive}, shape semantics modeling~\cite{li2017grass,li2019grains}, and document layout generation~\cite{patil2020read}. 
In contrast to previous data-driven methods, our recursive network fully exploits the hierarchical organization of the vessel and learns a low-dimensional manifold encoding branch connectivity along with geometry features describing the target surface. 
Once trained, the VesselVAE latent space is sampled to generate new vessel geometries. 
In this article we further expand our results and analysis. Specifically, we trained on different datasets of cerebral arteries, including aneurysms. We performed extensive experiments with our synthetically generated datasets to assess the ability of our method to generate high quality vessel geometry. These experiments build upon and enhance our previous research by providing more comprehensive results, thereby offering a deeper understanding of the capabilities of RvNNs in the generation of anatomically plausible vascular structures.

\begin{figure}[!t]
\includegraphics[width=\columnwidth]{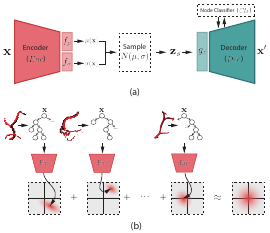}
\caption{(a) Overview of the Recursive variational Neural Network for synthesizing blood vessel structures. 
The architecture follows an Encoder-Decoder framework which can handle the hierarchical tree representation of the vessels. 
The encoder network ($Enc$) maps a training exemplar to a compact Gaussian distribution in the latent space, while the generator maps a sample from this Gaussian back to the original exemplar.
(b) The combination of per-exemplar Gaussians is restricted to match the standard normal distribution (bottom). 
In the testing phase, drawing samples from the standard normal and decoding them through the generator allows for exploration of the probabilistic manifold captured from the training set.
Note that vessel geometry is represented as a binary tree ($\mathbf{x}$) before being passed through the encoder.}
\label{fig:overview}
\end{figure}

\section{Methods}

\subsection{Input}
The network's  input\st{s} \pau{is} a binary tree representation of the blood vessel 3D geometry. 
Formally, each tree is defined as a tuple $(T, \mathcal{E})$, where $T$ is the set of nodes, and $\mathcal{E}$ is the set of directed edges connecting a pair of nodes ${(n, m)}$, with $ n,m \in T$.
In order to encode a 3D model into this representation, vessel segments are parameterized by a central axis consisting of ordered points in Euclidean space, each point has coordinates $x, y, z$ and radius $r$, assuming a piece-wise tubular vessel for simplicity. 
 We then construct the binary tree as a set of nodes $ T = {n_1, n_2, \ldots, n_N}$, where each node $n_i$ represents a vessel segment and contains an attribute vector 
$\mathbf{x}_i = [x_i, y_i, z_i, r_i] \in \mathbb{R}^4$ 
with the coordinates of the corresponding point and its radius $r_i$. See Sec.~\ref{sec:experimental} for details. \\

\subsection{Network architecture} 
The proposed generative model is a Recursive variational Neural Network (RvNN) trained to learn a probability distribution over a latent space that can be used to generate new blood vessel segments (Fig.~\ref{fig:overview}).
The framework follows an encoder-decoder structure, and the input is a binary tree representation of the vessel $\mathbf{x}$. During training, the encoder ($Enc$) maps this tree to a compact Gaussian distribution in the learned latent space, ensuring consistency with a standard normal distribution across all training exemplars.
After encoding, the decoder  ($Dec$) draws samples from a multivariate Gaussian distribution: $\mathbf{z}_s(\mathbf{x})$ $\sim N(\mu, \sigma)$, with $\mu=f_\mu(Enc((\mathbf{x}))$ and $\sigma=f_\sigma(Enc(\mathbf{x}))$, and learns to reconstruct the input vessel anatomies. After training, the RvNN can generate new geometry by drawing samples from a standard normal distribution and decoding them into novel blood vessel trees $\mathbf{x}'$.

\subsubsection{Encoder network} 
\begin{figure*}[ht]
\includegraphics[width=\textwidth]{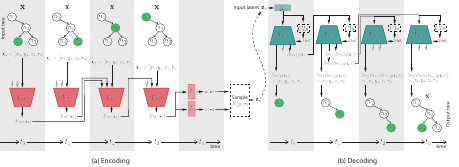}
\caption{Proposed encoder-decoder framework.
(a) Encoding. The input vessel tree is traversed over time steps ($t_0\!\rightarrow t_3$, encoding side) from leaves to root.
For each node (indicated in green), a feature vector $\mathbf{x}_i=[x_i,y_i,z_i,r_i]$ is processed by the encoder $Enc$ to produce latent embeddings that are ultimately merged into $\mu(\mathbf{x})$ and $\sigma(\mathbf{x})$ parameters of a Gaussian distribution in $t_4$.
(b) Decoding. A sample from this distribution is then fed through the decoder $Dec$ to hierarchically reconstruct the tree ($t_0\!\rightarrow t_3$, decoding side).
At each step, the model reconstructs output features for the node $\mathbf{x}_i=[x_i,y_i,z_i,r_i]$ and either continues branching or terminates (via the $Cls$ outputs), eventually reproducing the full vessel-like structure.  
Green circles highlight nodes created in each step.}
\label{fig:encodingdecoding}
\end{figure*}

The encoding process begins with a depth-first post order traversal of the tree $\mathbf{x}$, recursively encoding each node together with its child nodes, beginning from the leaf nodes until the root node is reached (Fig.~\ref{fig:encodingdecoding}, (a)).
First, the node’s attribute vector is processed to produce a partial encoding.
If the node has two children, their attribute vectors—along with those of any descendants—are already recursively encoded.
The network subsequently processes the latent vectors obtained from the right and left child nodes through additional layers, sums the resulting representations, and then concatenates them with the node’s partial encoding.
If the node has only one child, the same procedure is followed using a single latent vector. For leaf nodes, a zero vector of the appropriate dimension is concatenated in place of the missing child branches.
Ultimately, every node produces a $D$ dimensional encoding that captures the attributes of both the node and its entire sub-tree.
Repeating this process up the hierarchy yields a single latent vector representing the entire tree. A more detailed figure illustrating the encoder's internal structure is provided in the supplementary materials.
In addition to the core architecture, our Encoder is further augmented with two auxiliary shallow fully-connected neural networks: $f_\mu$ and $f_\sigma$. Positioned before the RvNN sampling step, these shallow networks shape the distribution of the latent space where the encoded tree structures lie.

\subsubsection{Decoder network} 
The decoding phase reverses the encoding process, reconstructing a binary tree representation $\mathbf{x}'$ from the sampled latent vector $\mathbf{z}_s$, as depicted in (Fig.~\ref{fig:encodingdecoding}, (b)).
The first step is to decode the root node. Then, on each recursive step, the network decides how many child nodes to decode (one, two, or none).
We equipped the decoder with
a node classifier module $Cls$, which predicts whether a given latent code corresponds to a leaf node or to an internal node with one or two bifurcations.
This is implemented as a multi-layer perceptron trained to predict a three-category bifurcation probability. Complementing the node classifier, the decoder layers are responsible for reconstructing the attribute vectors for each node, specifically its coordinates and radius. 
Additionally, the decoder can generate up to two extra latent codes (left and right) to enable the recursive decoding of the next node in the hierarchy.
If the classifier predicts that only one child is present, we assign it as the right child by default.
The Decoder is further augmented with an auxiliary shallow fully-connected neural network: $g_z$ situated after the sampling to facilitate the decoding of latent variables, aiding the Decoder network in the reconstruction of tree structures. Collectively, this supplementary network together with the ones of the Encoder ($f_\mu$ and $f_\sigma$) streamline the data transformation process through the model. All activation functions used in our networks are leaky ReLUs, except for $g_z$ which uses the Tanh. A complete description of the networks, including a table detailing layers, activation functions, and parameter counts, is provided in the supplementary materials.

\subsection{Objective} 
The encoder-decoder network and the node classifier are trained simultaneously in order to reconstruct the feature vectors $\mathbf{x}_i$ for each node along with the tree topology.
For this task, we propose a combined objective function consisting of \(L_{recon}\), which ensures feature vector reconstruction:

\begin{equation}
L_{recon} = \left\|Dec\left({g}(\mathbf{z}_s(\mathbf{x})\right))-\mathbf{x}\right\|_2.
\end{equation}
Note that $\mathbf{z}_s(\mathbf{x})$ denotes a sample drawn from the latent Gaussian distribution whose parameters are computed from the input $\mathbf{x}$ as described before. We also define a loss term to enforce topological correctness using a three-class cross-entropy loss:

\begin{equation}
L_{{topo}} = -\sum_{c=1}^{3} w_c x_{c} \log(p_{c}).
\end{equation}

 For each node, \(p_{c} = Cls(Dec\left({g_z}(\mathbf{z}_s(\mathbf{x})\right)))_c\) denotes the predicted probability that the node belongs to class \(c\). These probabilities are the output of the classifier, which operates on the features produced by the decoder. The classes represent different topological types of nodes: \textit{leaf nodes} (no children), \textit{internal nodes} (one child), and \textit{bifurcation nodes} (two children). The ground truth label \(x_{c}\) is a one-hot encoded vector that equals 1 if the true class of the node is \(c\), and 0 otherwise. To address class imbalance, each class is weighted by \(w_c\), the inverse of its frequency in the dataset.

Additionally, in line with the general framework proposed by {\mbox{$\beta$-VAE}}~\cite{higgins2017beta}, we incorporated a Kullback-Leibler (KL) divergence term encouraging the distribution $p(\mathbf{z}_s(\mathbf{x}))$ over all training samples $\mathbf{x}$ to move closer to the prior of the standard normal distribution $p(\mathbf{z})$. This regularization term is defined as:
\begin{equation}
L_{KL}=D_{KL}\left(p\left(\mathbf{z}_s(\mathbf{x})\right) \| p(\mathbf{z})\right).
\end{equation}
Finally, during training we minimize the following equation: 
\begin{equation}
    L = (1-\alpha) L_{recon} + \alpha L_{topo} + \gamma L_{KL}.
\label{eq:loss}
\end{equation}

\subsection{3D mesh synthesis} 

Several algorithms have been proposed to generate a mesh representation of a 3D surface from a tree-structured centerline~\cite{WU20134}.
We first converted the centerline to a Signed Distance Function (SDF) defined over the 3D space. We discretize the SDF by computing the shortest voxel-to-centerline distance. The SDF at a voxel position $\mathbf{v}$ $\in \mathbb{R}^3$ is defined as:

\begin{equation}
SDF(\mathbf{v}) = \min \limits_{\mathbf{c} \in C} \left( \|\mathbf{v} - \mathbf{c}\|^2 - r(\mathbf{c})^2 \right),
\end{equation}
where $C$ is the set of points along the centerline, $\mathbf{c}$ is a point on the centerline and $r(\mathbf{c})$ is the radius associated with $\mathbf{c}$. This equation computes the shortest distance from the voxel position $\mathbf{v}$ to the centerline $C,$ adjusted by radius $r(\mathbf{c})$.
We then apply the Marching Cubes algorithm~\cite{lorensen1998marching}, a widely used technique in computer graphics and computational geometry to convert volumetric data into triangle meshes. 
This process enabled us to preserve fine anatomical details and spatial relationships within the arterial network and obtain a representation suitable for further analysis.
Both the SDF generation and marching cubes algorithm were implemented using scripts from the VMTK toolkit~\footnote{\url{http://www.vmtk.org/documentation/vmtkscripts.html}} 

\section{Experimental Setup} \label{sec:experimental}

\subsection{Materials} 
We trained our networks on two different datasets. 
The first is a subset of the open-access Intra dataset~\footnote{\url{https://github.com/intra3d2019/IntrA}} published by Yang~et~al.~in 2020~\cite{yang2020intra}. 
This subset consisted of 100 healthy vessel segments reconstructed from 2D MRA images of patients.  
We also trained our architecture with the open-access Aneurisk dataset~\footnote{\url{https://github.com/permfl/AneuriskData}} published by the~\cite{AneuriskWeb}. This dataset is composed of 100 vessel segments reconstructed from 3D angiographic images containing healthy vessel segments and aneurysms. See Tab.~\ref{tab:node_stats} for details.

As a pre-process, we converted 3D meshes into a binary tree representation and used the \emph{network extraction} script from the VMTK toolkit~\footnote{\url{http://www.vmtk.org/vmtkscripts/vmtknetworkextraction}} to extract the centerline coordinates of each vessel model.
The centerline points were determined based on the ratio between the sphere step and the local maximum radius, which was computed using the advancement ratio specified by the user.
The radius of the blood vessel conduit at each centerline sample was determined using the computed cross-sections assuming a maximal circular shape (see Fig.~\ref{fig:dataset}).
To improve computational efficiency during recursive tree traversal, we implemented an algorithm that balances each tree by identifying a new root. 
The new root guarantees the lowest tree height possible without altering the represented vessel topology.
We experimented with trimming the trees to different heights and analyzed the network performance for each height. 
The decision to use lower tree heights reflects a balance between the computational demands of depth-first tree traversal in each training step and the complexity of the training meshes (see Fig.~\ref{fig:depth}). 
Note that it is still possible to train with deeper trees at the expense of higher computational costs.

Non-binary trees were transformed into binary trees through an iterative algorithm designed to split multifurcations into binary bifurcations. For each node with more than two children, the algorithm introduces intermediate nodes with the same position and features as the original. This intermediate nodes are then used to sequentially divide the children into pairs, ensuring that no node has more than two children. The process carefully preserves the original spatial arrangement and attributes of the vessel, altering only the tree topology. We excluded vessel models with loops from our experiments.
\\

\begin{figure}[!t]
\includegraphics[width=\columnwidth]{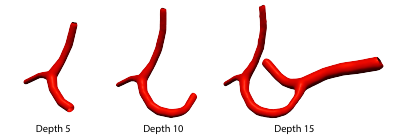}
\caption{Representation of tree structures at varying heights: The same vascular tree trimmed to different heights (5, 10, and 15), demonstrating the structural changes and complexity reduction at lower heights.}
\label{fig:depth}
\end{figure}

\begin{table}[h]
    \centering
    \begin{tabularx}{\columnwidth}{@{}ccXXXc@{}}
        \hline
        \multirow{2}{*}{Height} & \multirow{2}{*}{$\epsilon$} & \multicolumn{3}{c}{Nodes} & Bifurcations \\
        \cline{3-5}
         & &   Min & Max & Avg & Avg \\
        \hline
        \multicolumn{6}{c}{Aneurisk (\cite{AneuriskWeb})} \\
        \hline
        5  & 0.2 & 5 & 18 &10.01 & 1.86 \\
           & 0.1 & 5 & 18 & 9.65  & 1.53 \\
        10 & 0.2 & 12 &  43 & 25.00 & 3.38 \\
           & 0.1 & 10 & 46 & 23.02 & 2.62 \\
        15 & 0.2 & 22 & 73 & 41.56 & 4.80 \\
           & 0.1 & 15 & 73 & 38.44 & 3.73 \\
        \hline
        \multicolumn{6}{c}{Intra (\cite{yang2020intra})} \\
        \hline
        5  & 0.2 &  6 & 15  & 9.67  & 1.52 \\
           & 0.1 &  7 & 15  & 9.78  & 1.46 \\
        10 & 0.2 & 6 & 30  & 14.53 & 1.69 \\
           & 0.1 & 7 & 35  & 18.86 & 1.66 \\
        15 & 0.2 & 6 & 34  & 14.88 & 1.70 \\
           & 0.1 & 7 & 46  & 20.64 & 1.72 \\
        \hline
    \end{tabularx}
    \caption{Node and bifurcation statistics for the Aneurisk and Intra datasets, computed at different tree heights and resampling rates ($\epsilon$).}
    \label{tab:node_stats}
\end{table}

\subsection{Classifier training strategy} 

To calculate the $L_{recon}$ loss term, we bypassed the classifier's predictions and used the original tree topology for decoding during training.
This approach allows us to compare node attributes effectively, even if the classifier fails to make accurate topology predictions while being trained to predict the degree of each node. 
Given the imbalance in node distribution across classes, we adjust the weights in the cross-entropy loss calculation by using the inverse of each class count.
During preliminary experiments, we observed that accurately classifying nodes closer to the tree root is critical. 
A misclassification at the top levels (closer to the root) can lead to a cascading effect, impacting the reconstruction of all subsequent branches in the tree (i.e. skip reconstructing a branch). 
To account for this, we implemented and evaluated different weighting schemes assigning a weight to the cross-entropy loss based on the node depth and number of total sub-tree nodes. \\

\subsection{Implementation details} 
We developed the data preprocessing pipeline and network code using Python, VMTK and the PyTorch Framework.
For centerline extraction, we set the VMTK script's advancement ratio to $1.05$. 
This produces multiple cross-sections at bifurcations, where we select the one with the smallest radius to ensure alignment with the principal centerline direction. 
All tree attributes were normalized to the range $[0, 1]$.
In our training setup, the batch size was set to 4, and we used the ADAM optimizer with settings $\beta_1 = 0.9$, $\beta_2 = 0.999$, and a learning rate of $1 \times 10^{-4}$.
For Eq.~\ref{eq:loss}, we used $\alpha = 0.3$ and $\gamma = 0.001$. All models were trained for 20,000 epochs.

\begin{figure*}[t!]
\includegraphics[width=\textwidth]{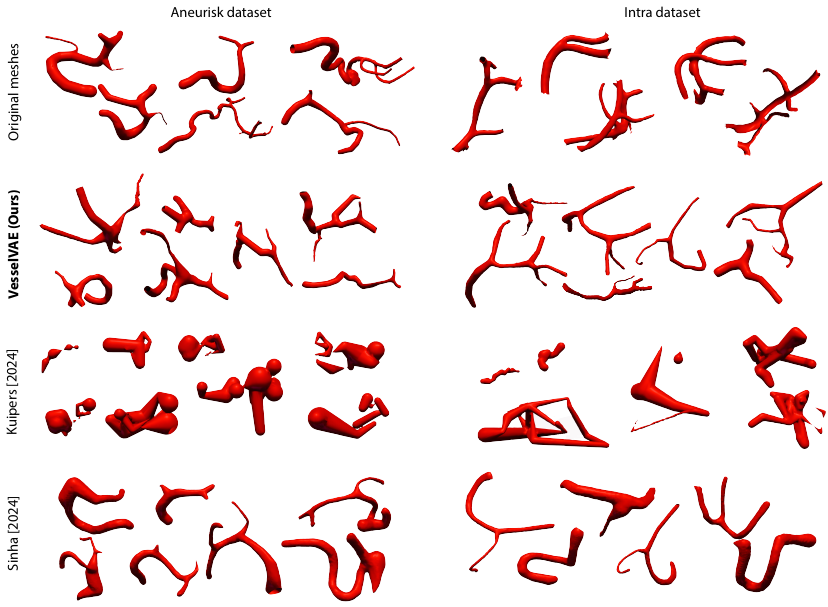}
\caption{Comparison of vessel geometries generated by VesselVAE (Ours) and baseline methods trained on the Aneurisk and Intra datasets. The approach by \cite{kuipers2024generating} struggles with intricate geometry, often producing disconnected segments and failing to capture the underlying feature distribution. Meanwhile, \cite{sinha2024representing} generates proper vessel geometries but lacks diversity, resulting in simpler and shallower structures. In contrast, our method demonstrates superior performance, producing diverse and realistic vessel meshes.}

\label{fig:results}
\end{figure*}
To accelerate computation, we implemented dynamic batching~\cite{looks2017deep}, which groups operations for input trees of varying shapes and different nodes within a single graph.
Training time is predominantly influenced by the height of the tree, as deeper trees require more sequential processing steps. Due to the implementation of dynamic batching, all nodes at the same depth are processed simultaneously, reducing the number of sequential steps required to transverse the tree. Consequently, while the amount of ramifications has some impact, it is relatively minor compared to the effect of tree height on training time. This is particularly noticeable with the Aneurisk dataset due to its larger number of ramifications compared to the  Intra dataset, see Tab.~\ref{tab:node_stats} for more details about the amount of bifurcations and nodes for each dataset.
On average, training our models takes between 12 and 36 hours.
All experiments were run on a workstation equipped with an NVIDIA A100 GPU, 80GB VRAM, and 256GB RAM.

\subsection{Experiments}
We trained our generative autoencoders using two different datasets, two different resampling rates, and three maximum heights to investigate how these factors and different hyperparameter choices influence the model's behavior and generative capabilities. 
After preprocessing and discarding vessels with loops, each dataset consisted of 100 samples.
We applied resampling to the original trees to better capture their structure with a reduced number of nodes. This was achieved using the Ramer-Douglas-Peucker algorithm~\cite{ramer1972iterative, douglas1973algorithms}. Although resampling can affect the height of the original tree, we imposed a fixed maximum during training (5, 10, or 15) to ensure consistency across experiments. The resampling rate, defined by parameters $\epsilon = 0.1$ and $\epsilon = 0.2$, controls the maximum allowable distance between the original and the simplified curve, influencing how well the tree's geometry is captured within the specified node limit.
Additionally, we conducted an ablation study to assess the impact of the weighting scheme on the node classifier behavior.

\section{Results}

We now evaluate our method against recent approaches (Sec.~\ref{sec:baseline_comparison}), beginning with a comparison with the methods of \cite{sinha2024representing} and \cite{kuipers2024generating} in Sect.~\ref{sec:sinha-kui}. This is followed by an analysis of results relative to the older baselines proposed by \cite{hamarneh2010vascusynth} and \cite{wolterink2018blood} in Sec.~\ref{sec:vascu-wolt}. In Sec.~\ref{sec:ane}, we further explore the synthesis of pathological geometry features within vascular structures. Lastly, Sec.~\ref{sec:ablation} presents an ablation study to evaluate the impact of relevant training configuration choices.

\definecolor{verylightgray}{gray}{0.88} 
\definecolor{lightgray}{gray}{0.95}

\begin{table}[t]
    \centering
    \begin{tabularx}{\columnwidth}{lccc}
        \hline
         & MMD ↓ & COV ↑ & 1-NNA \\
        \cmidrule{2-4}
        & \multicolumn{3}{c}{Aneurisk} \\
        \cmidrule{1-4}
        
        \cite{sinha2024representing} & 1.264 & .19 & .03 \\
        \cite{kuipers2024generating} &  .537 & .03 & .0 \\
        \textbf{VesselVAE (Ours)} & \textbf{.004} & \textbf{.58} & \textbf{.68} \\
        \cmidrule{1-4}
        & \multicolumn{3}{c}{Intra} \\
        \cmidrule{1-4}
        
        \cite{sinha2024representing} & .447 & .22 & .1\\
        \cite{kuipers2024generating} & .288 & .04 & .0 \\
        \textbf{VesselVAE (Ours)} & \textbf{.039} & \textbf{.47} & \textbf{.4} \\
        \hline
    \end{tabularx}
    \caption{Quantitative results for novel tree generation comparing our method against baseline approaches. Metrics are denoted as follows: ↑ indicates that higher values are better, ↓ indicates that lower values are better, and for 1-NNA, the optimal score is 0.5. For \cite{sinha2024representing} and \cite{kuipers2024generating}, we observed low 1-NNA alongside low COV, indicating model collapse: the model produces highly similar samples that cluster in a narrow region of the distribution, sufficiently realistic to fool the 1-NN classifier, but lacking diversity.}
    \label{tab:evaluation}
\end{table}
\subsection{Comparison against baseline methods.} \label{sec:baseline_comparison}
\subsubsection{\cite{sinha2024representing} and \cite{kuipers2024generating}} \label{sec:sinha-kui}

We quantitatively evaluated the generated samples and compared them to recent methods using three established metrics: Minimum Matching Distance (MMD), Coverage (COV), and 1-Nearest Neighbor Accuracy (1-NNA).

MMD~\cite{achlioptas2018learning} measures the quality of the generated samples by computing, for each real sample, the distance to its closest counterpart in the generated set and averaging these values. Low MMD values indicate that the generated samples are close to the real ones, suggesting high fidelity and good coverage of the data distribution. COV~\cite{achlioptas2018learning} quantifies the diversity of the generated distribution by determining the proportion of real samples that can be paired with at least one generated sample. This is done by identifying the closest match in the reference set for each generated sample using a distance function such as Chamfer Distance (CD) or Earth Mover’s Distance (EMD). Higher COV values are desirable, as they indicate that a larger portion of the real distribution is represented by the generated set. Although coverage is effective in detecting mode collapse (e.g. when many generated samples map to the same or a few real samples), it does not provide information about sample quality. In some cases, perfect coverage can be achieved even if the matched pairs are significantly different. Finally, 1-NNA~\cite{yang2019pointflow} evaluates the plausibility of the generated distribution by computing the leave-one-out 1-nearest neighbor classification accuracy between real and generated samples. A value close to 50\% indicates that the two distributions are well-mixed, while scores significantly above or below this threshold suggest mode collapse or poor generalization.

Our evaluation was performed on both datasets, resampled at a rate of $0.2$ and with a depth of $10$, as this configuration yielded the best performance in our ablation study.
The results are presented in Tab.~\ref{tab:evaluation}. 
Our model consistently outperforms both baselines across all three metrics when trained on the same datasets. Additionally, Fig.\ref{fig:results} complements the quantitative analysis with example renders for all methods and training datasets.
In terms of efficiency, our method is significantly more lightweight, using $\sim$1 million parameters compared to the $\sim$43 million and $\sim$41 million parameters required by \cite{sinha2024representing} and \cite{kuipers2024generating}, respectively. 
Furthermore, our model has a much smaller memory footprint during training, requiring $\sim$67MB compared to $\sim$173MB and $\sim$2,000MB for the other methods.

\subsubsection{\cite{hamarneh2010vascusynth}} \label{sec:vascu-wolt}
Comparing synthetic samples with real datasets (such as Aneurisk or Intra) is possible by measuring differences in parameter distributions. 
However, it is not trivial to compare our approach with other methods such as Vascusynth.
This is because procedural and data-driven methods have fundamentally different approaches, making direct comparison challenging.

To make this feasible, we generated a synthetic dataset using the method described in \cite{hamarneh2010vascusynth} and trained VesselVAE on these data. 
The goal is to assess whether VesselVAE can generate similar samples while maintaining the expressiveness and precision of the original generative method.
In our experiments, the training dataset was resampled with $\epsilon=$ 0.1 and $\epsilon=$ 0.2 (as with  previous comparisons).
After training, we computed radius, tortuosity, and length metrics to compare distributions.
Table summarizing results can be found in Supplementary Material.
Fig.~\ref{fig:resultsvascu} illustrates vessel geometries generated by VesselVAE when trained on Vascusynth data, demonstrating its ability to capture the main geometric characteristics of Vascusynth.

\subsection{Pathology geometry synthesis} \label{sec:ane}
\begin{figure}[!t]
\includegraphics[width=\columnwidth]{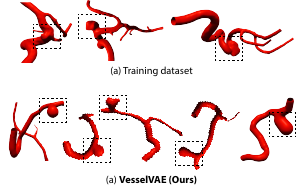}
\caption{(a) Vessels with aneurysms from the Aneurisk dataset. (b) Vessels with aneurysms generated by VesselVAE.}
\label{fig:ane}
\end{figure}

A key strength of our data-driven approach is its ability to naturally capture and reproduce distinctive geometric features present in the training data. In Fig.~\ref{fig:ane} we show some synthesized vessel geometries featuring aneurysms after trainig on the Aneurisk dataset. Unlike parametric methods which require explicit coding to replicate such features, our approach inherently learns and reproduces both realistic non-pathological and pathological vessel structures.

\begin{figure}[t!]
\includegraphics[width=\columnwidth]{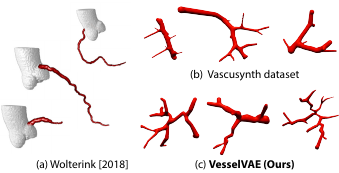}
\caption{(a) Vessel geometries generated by \cite{wolterink2018blood}. The evaluation is limited to a visual comparison due to the lack of access to the original code or data. (b) Vessel geometries generated using the method presented in \cite{hamarneh2010vascusynth}. (c) Vessel geometries generated by our method when trained on (b). While occasionally struggling with unrealistic angles, VesselVAE effectively captures the underlying characteristics of the data, demonstrating a similar expressive power to the parametric method proposed by \cite{hamarneh2010vascusynth}.}
\label{fig:resultsvascu}
\end{figure}

\begin{table*}[!t]
  \centering
  \footnotesize
     \begin{tabularx}{\textwidth}{>{\hspace{0pt}}c<{\hspace{0pt}}cccccccccccccccc@{}}
       \hline
       
       & \multicolumn{2}{c}{} & \multicolumn{6}{c}{Cosine Similarity ↑}  & & \multicolumn{6}{c}{Earth Mover Distance ↓}\\
       
        \cmidrule{4-9} \cmidrule{11-16} 
       \multicolumn{3}{c}{} & \multicolumn{2}{c}{Radius} & \multicolumn{2}{c}{Tortuosity} & \multicolumn{2}{c}{Length} && \multicolumn{2}{c}{Radius}  &\multicolumn{2}{c}{Tortuosity} & \multicolumn{2}{c}{Length} \\
       
       \multirow{2}{*}{\makecell{Weighting \\ Scheme}} & \multirow{2}{*}{\makecell{Max. \\ Height}} & \multirow{2}{*}{$\epsilon$} &\multirow{2}{*}{M}  &\multirow{2}{*}{CI}  &\multirow{2}{*}{M} & \multirow{2}{*}{CI}  &\multirow{2}{*}{M} & \multirow{2}{*}{CI} && \multirow{2}{*}{M} & \multirow{2}{*}{CI} & \multirow{2}{*}{M} & \multirow{2}{*}{CI}  &\multirow{2}{*}{M} & \multirow{2}{*}{CI}\\
       \\
       \hline
       &\multicolumn{3}{c}{} & \multicolumn{11}{c}{Aneurisk(~\cite{AneuriskWeb}} \\
       \hline
        \emph{Uniform} & 5 & 0.2 & .93 & (.91 .95)  & .88 & (.86 .89) & .95 & (.93 .97) && .55 & (.47 .54)  & 2.53 & (1.99 3.06) & .65 & (.48 .82) \\
       \rowcolor{verylightgray} \emph{Depth} & &  & .95 & (.93 .96) & .92 & (.90 .94) & .96 & (.95 .97)   && .39 & (.28 .49)  & 1.87 & (1.44 2.29) & .45 & (.28 .61) \\
       \rowcolor{lightgray} \emph{Sub-tree} &  & &.94 & (.92 .96)  & .93 & (.91 .94) & .96 & (.93 .98) && .43 & (.33 .52)  & 2.24 & (1.81 2.66) & .28 & (.24 .32) \\
         \emph{Uniform} &  & 0.1 & .94 & (.93 .96)   & .77 & (.72 .83)& .93 & (.91 .95) && .49 & (.35 .62)   & 3.1 & (2.42 3.79)& .64 & (.47 .8) \\
        \rowcolor{verylightgray} \emph{Depth}&&  & .92 & (.90 .94) & .83 & (.79 .86) & .96 & (.94 .97) && .54 & (.4 .68) & 2.49 & (2.13 2.86)& .43 & (.33 .53)  \\
       
        \rowcolor{lightgray} \emph{Sub-tree} &  & &.92 & (.91 .94)  & .9 & (.87 .92) & .95 & (.95 .96) && .54 & (.41 .66)  & 2.61 & (1.89 3.34) & .39 & (.28 .5) \\
       \hline
        \emph{Uniform}&10& 0.2 & .95 & (.93 .96)   & .86 & (.83 .89)& .92 & (.9 .94) && .68 & (.54 .82)   & 2.37 & (1.99 2.75) & .82 & (.67 .97) \\
       \rowcolor{verylightgray} \emph{Depth}& &  & .93 & (.90 .95)  & .93 & (.91 .96) & .96 & (.94 .97) && .58 & (.44 .72)  & 1.21 & (1.08 1.34) & .38 & (.22 .54) \\
       
       \rowcolor{lightgray} \textbf{\emph{Sub-tree}} &  & & \textbf{.96} & \textbf{(.94 .98)}  & \textbf{.94} & \textbf{(.92 .95)} & \textbf{.96} & \textbf{(.95 .97)} && \textbf{.58} & \textbf{(.47 .69)}  & \textbf{1.15} & \textbf{(.85 1.45)} & \textbf{.36} & \textbf{(.28 .44)} \\
        \emph{Uniform}& & 0.1 & .91 & (.9 .93)   & .77 & (.7 .83)& .86 & (.82 .9) && .74 & (.55 .94)   & 3.08 & (2.59 3.56)& 1.27 & (1.12 1.42) \\
        \rowcolor{verylightgray} \emph{Depth}& & & .94 & (.93 .95)  & .91 & (.88 .95)& .96 & (.95 .97) && .49 & (.42 .56)  & 1.5 & (1.22 1.79)& .59 & (.46 .71) \\
        
        \rowcolor{lightgray} \emph{Sub-tree} &  & & .96 & (.95 .97)  & .93 & (.9 .95) & .95 & (.93 .97) && .47 & (.36 .58)  & 1.6 & (1.15 2.05) & .46 & (.29 .63) \\
       \hline
        \emph{Uniform} & 15 & 0.2 & .83 & (.8 .86) & .83 & (.81 .85)& .91 & (.88 .93)   && 1.25 & (1.14 1.31)   & 1.92 & (1.37 2.48)& 1.4 & (1.18 1.62) \\
        \rowcolor{verylightgray} \emph{Depth} & &  & .92 & (.9 .94)  & .83 & (.78 .87) & .46 & (.4 .51) && .6 & (.47 .72)  & 2.64 & (1.87 3.4) & 2.72 & (2.54 2.91)\\
       
       \rowcolor{lightgray} \emph{Sub-tree} & &  & .92 & (.9 .95)  & .95 & (.94 .96) & .94 & (.94 .95) && .54 & (.42 .65)  & 1.19 & (.96 1.41) & .81 & (.71 .91) \\
        \emph{Uniform} & & 0.1 & .94 & (.92 .96)   & .74 & (.68 .8)& .86 & (.84 .88) && .76 & (.64 .88)   & 3.52 & (2.82 4.22)& 1.6 & (1.44 1.76) \\
        \rowcolor{verylightgray} \emph{Depth} & &  & .93 & (.92 .95)  & .84 & (.8 .88)& .72 & (.69 .76) && .63 & (.52 .73)  & 2.13 & (1.55 2.71)& 1.91 & (1.8 2.02) \\
        
        \rowcolor{lightgray} \emph{Sub-tree} &  & & .94 & (.92 .95)  & .83 & (.81 .86) & .91 & (.88 .94) && .77 & (.65 .88)  & 1.79 & (1.32 2.26) & .75 & (.64 .85) \\
       \hline
       
       &\multicolumn{3}{c}{} & \multicolumn{11}{c}{Intra(~\cite{yang2020intra}} 
       
       \\
       \hline
       \emph{Uniform}  & 5 & 0.2 & .97 & (.96 .98)  & .93 & (.92 .94) & .93 & (.9 .95) && .55 & (.42 .67)  & 2.38 & (1.79 2.97) & .48 & (.4 .55) \\
        \rowcolor{verylightgray} \emph{Depth} & &  & .95 & (.94 .96) & .93 & (.91 .95) & .93 & (.91 .95)   && .56 & (.46 .65)  & 1.7 & (1.3 2.09) & .39 & (.32 .46) \\
       
       \rowcolor{lightgray} \emph{Sub-tree} &  & & .94 & (.92 .96)  & .94 & (.91 .96) & .94 & (.92 .96) && .42 & (.29 .56)  & 1.86 & (1.41 2.32) & .52 & (.42 .63) \\
        \emph{Uniform}&& 0.1 & .97 & (.96 .98)   & .91 & (.89 .95)& .95 & (.93 .96) && .44 & (.35 .53)   & 2.7 & (2.08 3.32)& .42 & (.35 .5) \\
        \rowcolor{verylightgray} \emph{Depth}& &  & .9 & (.88 .91) & .9 & (.87 .94) & .96 & (.95 .97) && .59 & (.53 .65) & 2.64 & (1.96 3.31)& .52 & (.4 .64)  \\
        
        \rowcolor{lightgray} \emph{Sub-tree} &  & &.97 & (.96 .98)  & .9 & (.87 .93) & .95 & (.93 .97) && .51 & (.42 .6)  & 3.49 & (2.3 4.69) & .39 & (.33 .45) \\
       \hline
        \emph{Uniform} & 10 & 0.2 & .95 & (.94 .97)   & .89 & (.87 .92)& .93 & (.9 .96) && .66 & (.52 .79)   & 2.33 & (1.66 3.0) & .41 & (.33 .49) \\
       \rowcolor{verylightgray} \textbf{\emph{Depth}}& &  & \textbf{.97} & \textbf{(.95 .99)} & \textbf{.97} & \textbf{(.95 .98)} & \textbf{.95} & \textbf{(.94 .97)} && \textbf{.41} & \textbf{(.37 .45)}  & \textbf{1.52} & \textbf{(1.06 1.98)} & \textbf{.42} & \textbf{(.32 .53)} \\
       
       \rowcolor{lightgray} \emph{Sub-tree} &  & & .95 & (.93 .97)  & .93 & (.9 .95) & .96 & (.93 .98) && .52 & (.42 .61)  & 1.6 & (1.32 1.88) & .31 & (.23 .39) \\
         \emph{Uniform} &  & 0.1 & .96 & (.94 .97)   & .81 & (.77 .85)& .95 & (.94 .97) && .39 & (.34 .43)   & 2.81 & (1.85 3.78)& .68 & (.61 .76) \\
        \rowcolor{verylightgray} \emph{Depth}&&  & .94 & (.93 .96)  & .95 & (.93 .97)& .94 & (.92 .96) && .56 & (.4 .73)  & 1.39 & (1.0 1.78)& .68 & (.55 .8) \\
       
        \rowcolor{lightgray} \emph{Sub-tree} &  & & .94 & (.92 .96)  & .86 & (.81 .9) & .95 & (.92 .97) && .54 & (.43 .66)  & 2.04 & (1.74 2.34) & .41 & (.25 .56) \\
       \hline
         \emph{Uniform} & 15 & 0.2 & .94 & (.92 .96) & .9 & (.88 .93)& .92 & (.89 .94)   && .67 & (.51 .83)   & 2.11 & (1.59 2.62)& .54 & (.39 .69) \\
        \rowcolor{verylightgray} \emph{Depth}& &  & .95 & (.93 .97)  & .96 & (.94 .97) & .92 & (.9 .94) && .43 & (.36 .49)  & 1.08 & (.84 1.31) & .48 & (.37 .59)\\
      
       \rowcolor{lightgray} \emph{Sub-tree} &  & & .96 & (.95 .97)  & .94 & (.91 .96) & .96 & (.95 .98) && .38 & (.29 .47)  & 1.63 & (1.33 1.93) & .38 & (.29 .47) \\
       \emph{Uniform} & & 0.1 & .95 & (.94 .97)   & .88 & (.83 .93)& .95 & (.93 .97) && .65 & (.53 .78)   & 2.71 & (1.85 3.57)& .93 & (.75 1.11) \\
         \rowcolor{verylightgray} \emph{Depth}& &  & .96 & (.93 .98)  & .93 & (.9 .96)& .95 & (.94 .97) && .54 & (.44 .65)  & 2.6 & (2.14 3.06)& 1.01 & (.82 1.19) \\
        
        \rowcolor{lightgray} \emph{Sub-tree} &  & & .94 & (.92 .96)  & .86 & (.81 .91) & .95 & (.93 .96) && .67 & (.56 .79)  & 1.64 & (1.14 2.14) & .56 & (.42 .69) \\
       \hline
   \end{tabularx}
  \caption{Ablation summary. Cosine Similarity and Earth Mover Distance for all the evaluated metrics with resampling parameter $\epsilon$ set to 0.1 and 0.2. All experiments were carried out 10 times with randomized seeds. We report the mean values (M) and 95 \% confidence intervals (CI).  The weighting scheme column corresponds to loss weight ablation. \emph{Uniform} refers to constant weighting, \emph{Depth} indicates depth-based weighting, and \emph{Sub-tree} corresponds to sub-tree size weighting. In bold the best-performing scheme for trees of depth 10 and resampling rate 0.2. These highlighted models were used for the evaluation in Tab.~\ref{tab:evaluation} and Fig.~\ref{fig:results}}
  \label{tab:resultados}
\end{table*}

\subsection{Ablation} \label{sec:ablation}
We conducted an ablation study to evaluate the impact of various configurations on our model's performance. Specifically, we tested three different loss weighting schemes: (1) assigning a \emph{uniform} weight of one to all nodes in the tree, (2) weighting each node based on its \emph{depth} within the tree, and (3) weighting each node according to the size of its \emph{sub-tree}, where each node is treated as the root of its respective sub-tree. Additionally, we explored the effects of two different vessel centerline resampling rates $\epsilon = 0.1$ and $\epsilon = 0.2$ and retrained our models using three different maximum heights (5, 10, and 15). This comprehensive evaluation allowed us to analyze the contributions of each factor to the overall performance of the model and select the best-performing configuration for the aforementioned comparisons.

To asses the best-performing model we established a set of metrics: tortuosity per branch, total vessel centerline length, and average tree radius.
These metrics have been widely used in the field of 3D modeling of blood vessels, and have been shown to provide reliable and accurate quantification of blood vessels' main characteristics~\cite{bullitt2003vascular,lang2012three, sinha2024representing}.
The tortuosity distance metric~\cite{bullitt2003measuring}, a measure of branch twistiness computed as the ratio between each branch's length and the straight-line distance between its endpoints, has clinical importance in blood vessel analysis. 
We also evaluated total vessel length and average radius metrics previously used to differentiate healthy from cancerous malformations~\cite{bullitt2003vascular,lang2012three}. 
Due to the inherent randomness in the generation process, we synthesized 10 distinct sets of 100 samples each for every combination of weights, dataset, depth, and resampling rate.
Metrics were computed for each training and generated set, and we compared distributions using Cosine Similarity (CS) and Earth Mover's Distance (EMD). We report the mean values along with 95~\% confidence intervals in Tab.~\ref{tab:resultados}.

\begin{figure}[t!]
\includegraphics [width=\columnwidth]{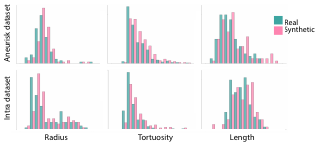}
\caption{Quantitative evaluation of generated vessel samples. The histograms display the distributions of each evaluated metric. Data generated by VesselVAE is labeled as \emph{synthetic} (pink), while data from the datasets is labeled as \emph{real} (blue). All histograms are computed from vessels with maximum height 10 and resampling $\epsilon$ 0.2}
\label{fig:histogramas}
\end{figure}

We observed high similarity scores between histograms at depths 5 and 10, indicating that the generated blood vessels are realistic. The slight decrease in similarity at depth 15, particularly for the Aneurisk dataset, is attributed to increased computational demands (see Section~\ref{sec:discussion}).
The  uniform scheme significantly deteriorates the similarity values across all evaluated metrics.
The effect of non uniform weighting gains importance when the height of the tree increases, while
for shallow trees (height 5 in our experiments) performance remains mostly unaffected.
Fig. ~\ref{fig:histogramas} shows the distribution for the evaluated metrics of the best-performing models for trees of maximum height 10.

To identify the best-performing scheme, we computed the average Earth Mover's Distance (EMD) across all metrics—Radius, Tortuosity, and Length—for each row in Table~\ref{tab:resultados}. We then averaged the EMD for each weighting scheme, resulting in a single EMD value per scheme (Intra: \emph{Uniform}=0.928, \emph{Depth}=0.938, \emph{Sub-tree}=0.933; Aneurisk: \emph{Uniform}=0.87, \emph{Depth}=0.87, \emph{Sub-tree}=0.93). In the Aneurisk dataset, the sub-tree size weighting scheme exhibited superior performance, while in the Intra dataset, the depth-based weighting scheme achieved slightly better results compared to the others. This variation is likely attributable to the greater number of ramifications in the Aneurisk dataset (see Tab.~\ref{tab:node_stats} for dataset statistics), which has a stronger influence on the computation of loss weights. Conversely, the Intra dataset, with its fewer ramifications, showed comparable performance across both schemes.
This highlights the role of this weight parameter in the autoencoder's capability to accurately synthesize vasculature geometry.

\section{Discussion and Limitations}
\label{sec:discussion}
\begin{figure}[!t]
\includegraphics[width=\columnwidth]{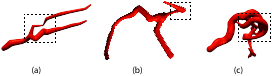}
\caption{Examples of VesselVAE generations exhibiting implausible angles (a, b) and self-intersections (c).}
\label{fig:fallas}
\end{figure}

Our experiments utilized two open-source datasets of cerebral vasculature, resampled at different rates to show the impact on evaluation metrics. Results indicate that complexity increases with tree height, leading to good results with lower tree depths and suboptimal outcomes at the highest depths tested.
While we experimentally found an optimal bottleneck size in our experiments, we hypothesize that the recursive encoding framework may lose information with high hierarchies. A per-branch evaluation could offer deeper insights (as proposed by Kuipers et al.). However, it requires labeled data which is not available in our setting.
Additionally, the generated geometries might show self-intersections. This issue arises because, although the model effectively generates plausible geometries, it does not enforce strict constraints to eliminate self-intersections.
In the future, we would like to incorporate restrictions into the generative model to avoid such artifacts.
Furthermore, while our method synthesizes vessels highly similar to those in the training dataset, some exhibit implausible angles. This suggests the model struggles with angle learning. Potential improvements include adding an attribute for angles, exploring different angle representations, or resampling arteries with greater detail on curves. Examples of failed cases can be seen in Fig.~\ref{fig:fallas}.

\section{Conclusions}
We have presented a novel approach for synthesizing blood vessel models using a Recursive variational
Neural Network.
Our method enables efficient encoding and decoding of binary tree structures and produces high-quality synthesized models. 
Our findings highlight the versatility of our network architecture, demonstrating its ability to synthesize blood vessels across different datasets, including those with aneurysms. Additionally, our approach outperforms recent state-of-the-art methods by more accurately replicating intricate vascular characteristics while producing more realistic and diverse geometries.
In the future, we aim to explore other surface representations e.g. the zero level set in a differentiable neural implicit (INR) \cite{sinha2024representing, alblas2023going}. 
This could lead to more accurate and efficient modeling of blood vessels and potentially other non-tree-like structures such as capillary networks. 
Another potential direction is the development of an autoregressive model utilizing a tokenized representation of the tree, enabling coherent generation and supporting conditional outputs.
Overall, our proposed approach shows potential for improving 3D blood vessel geometry synthesis and supporting the development of clinical tools that could benefit healthcare professionals.
\section*{Acknowledgments}
This project was supported by grants from Salesforce, USA (Einstein AI 2020), National Scientific and Technical Research Council (CONICET), Argentina (PIP 2021-2023 GI - 11220200102981CO), and Universidad Torcuato Di Tella, Argentina.

\section*{Conflict of interest statement}
The authors have no conflicts to disclose.

\bibliographystyle{model2-names.bst}\biboptions{authoryear}
\bibliography{bibliography}
\clearpage

\renewcommand{\thefigure}{\arabic{figure}s}
\setcounter{figure}{0}

\renewcommand{\thetable}{\arabic{table}s}
\setcounter{table}{0}

\section*{Supplementary Material}
\begin{table}[!h]
\centering

\label{tab:architecture}
    \begin{tabular}{@{}ccccccc@{}}
\hline
Layer name & Input shape & Output shape & \phantom{abc} & Layer name & Input shape & Output shape   \\\cmidrule{1-3} \cmidrule{5-7}
\multicolumn{3}{c}{\textbf{Encoder}} && \multicolumn{3}{c}{\textbf{Decoder}} \\
\cmidrule{1-3} \cmidrule{5-7}
fc1 & 4 & 512 && fc1 & 64 & 256 \\
fc2 & 512 & 64 && fc\_left1 & 256 & 256 \\
right\_fc1 & 64 & 512 && fc\_left2 & 256 & 64 \\
right\_fc2 & 512 & 64 && fc\_right1 & 256 & 256 \\
left\_fc1 & 64 & 512 && fc\_right2 & 256 & 64 \\
left\_fc2 & 512 & 64 && fc2 & 256 & 64 \\
fc3 & 128 & 64 && fc3 & 64 & 4 \\
\hline
\multicolumn{7}{c}{ Supplementary Networks}  \\ 
\hline
\multicolumn{3}{c}{ $f_\mu$ + $f_\sigma$} & \phantom{abc} & \multicolumn{3}{c}{$g_z$} \\ \cmidrule{1-3} \cmidrule{5-7}
fc1 & 128 & 512 && fc1 & 64 & 256 \\ 
fc2mu & 512 & 128 && fc2 & 256 & 256 \\ 
fc2var & 512 & 128 && fc3 &  256 & 64 \\ \cmidrule{1-3} \cmidrule{5-7}
\multicolumn{3}{c}{\;} & \phantom{abc}& \multicolumn{3}{c}{Classifier} \\  \cmidrule{5-7}
- & - & - && fc1 & 64 & 256 \\ 
- & - & - && fc2 & 256 & 256 \\ 
- & - & - && fc3 & 256 & 3 \\ 
\hline

\end{tabular}
\caption{Our proposed method consists of two primary components: a recursive Encoder and a recursive Decoder. All layers in each component are fully connected with leaky rectified linear unit (Leaky ReLU) activations, except for $g_z$ which uses the Tanh. The Classifier is a sub-network inside the Decoder that predicts a label for each node in the input graph.}
\end{table}

\begin{table}[h]
    \centering
    \begin{tabular}{lccccc}
        \hline
        & Encoder & Decoder & $f_\mu$ + $f_\sigma$ & $g_z$ & Classifier \\
        \hline
        Trainable parameters & 527616 & 197828 & 98944 & 9880 & 83203\\
        \hline
    \end{tabular}
    \caption{Summary of trainable parameters for each component.}
    \label{tab:trainable_params}
\end{table}

\begin{figure*}[b]
\includegraphics[width=\textwidth]{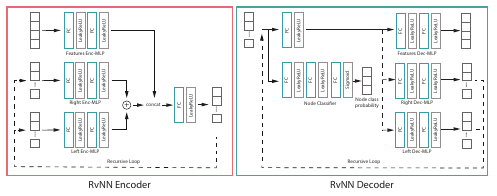}
\caption{Layers of the Encoder and Decoder networks comprising
branches of fully-connected layers followed by leaky ReLU activations. Notice that
right/left Enc-MLPs of the Encoder only execute when the incoming tree requires it.
Similarly, the Decoder only uses right/left Dec-MLPs when the Node Classifier predicts
bifurcations.}
\label{fig:depth}
\end{figure*}

\begin{table*}[t]
    \centering
    \footnotesize
    \begin{tabularx}{\textwidth}{>{\hspace{0pt}}c<{\hspace{0pt}}cccccccccccccccc@{}}
        \toprule
        \multicolumn{3}{c}{} & \multicolumn{13}{c}{Vascusynth}  \\
        \cmidrule{4-16}
         \multicolumn{3}{c}{} & \multicolumn{6}{c}{Cosine similarity ↑}  & & \multicolumn{6}{c}{Earth Mover Distance ↓}\\
        \cmidrule{4-9}\cmidrule{11-16}
        \multicolumn{3}{c}{} & \multicolumn{2}{c}{Radius} & \multicolumn{2}{c}{Tortuosity} & \multicolumn{2}{c}{Length} && \multicolumn{2}{c}{Radius}  &\multicolumn{2}{c}{Tortuosity} & \multicolumn{2}{c}{Length} \\
        \cmidrule{4-9}\cmidrule{11-16}
        
        \multirow{2}{*}{\makecell{Weighting \\ Scheme}} & \multirow{2}{*}{\makecell{Max. \\ Height}} & \multirow{2}{*}{$\epsilon$} &\multirow{2}{*}{M}  &\multirow{2}{*}{CI}  &\multirow{2}{*}{M} & \multirow{2}{*}{CI}  &\multirow{2}{*}{M} & \multirow{2}{*}{CI} && \multirow{2}{*}{M} & \multirow{2}{*}{CI} & \multirow{2}{*}{M} & \multirow{2}{*}{CI}  &\multirow{2}{*}{M} & \multirow{2}{*}{CI}\\
        \\
        \cline{3-8}
         
        \hline
        \emph{Depth} &5 & 0.2 & .96 & (.95 .97) & .51 & (.47 .56) & .80 & (.74 .85) & & .42 & (.32 .53) & 2.67 & (1.83 3.52) & .44 & .3 .59\\
        \rowcolor{lightgray} \emph{Sub-tree} & &  & .92 & (.89 .95)  & .78 & (.74 .83) & .93 & (.91 .95) && .54 & (.45 .64)  & 3.35 & (2.68 4.01) & .46 & (.33 .58)\\
         \emph{Depth}&& 0.1 & .92 & (.89 .94) & .87 & (.83 .91) & .74 & (.64 .83) & & .41 & (.3 .52) & 3.64 & (2.85 4.43) & 1.05 & (.91 1.19)\\
         \rowcolor{lightgray} \emph{Sub-tree} &&  & .9 & (.88 .92)  & .6 & (.55 .64) & .85 & (.82 .87) && .69 & (.52 .85)  & 5.4 & (4.33 6.47) & 1.77 & (1.59 1.94)\\
        \hline
        \emph{Depth}&10 & 0.2 & .87 & (.85 .89) & .93 & (.91 .96) & .88 & (.85 .9) & & .61 & (.42 .8) & 1.95 & (1.43 2.47) & .46 & (.37 .55)\\
        \rowcolor{lightgray} \emph{Sub-tree}&&  & .91 & (.88 .92)  & .85 & (.83 .86) & .96 & (.95 .97) && .97 & (.7 1.23)  & 1.18 & (.93 1.44) & .85 & (.71 .98)\\
         \emph{Depth}&& 0.1 & .96 & (.95 .97) & .91 & (.89 .93) & .83 & (.81 .85) & & .5 & (.38 .62) & 2.68 & (1.86 3.49) & .66 & (.54 .79)\\
         \rowcolor{lightgray} \emph{Sub-tree}&&  & .89 & (.88 .9)  & .63 & (.55 .71) & .82 & (.79 .84) && 1.0 & (0.9 1.11)  & 3.33 & (1.91 4.76) & 1.57 & (1.42 1.72)\\
        \hline
        \emph{Depth}&15 & 0.2 & .5 & (.47 .53) & .65 & (.59 .72) & .56 & (.49 .63) & & 2.65 & (2.34 2.96) & 1.38 & (1.02 1.74) & 2.9 & (2.53 3.26)\\
        \rowcolor{lightgray} \emph{Sub-tree}& &  & .82 & (.79 .85)  & .76 & (.68 .84) & .87 & (.85 .89) && 1.4 & (1.28 1.53)  & 2.41 & (1.51 3.3) & .77 & (.64 .89)\\
        \emph{Depth}& & 0.1 & .86 & (.82 .9) & .82 & (.79 .85) & .74 & (.68 .79) & & .87 & (.79 .94) & 1.41 & (1.02 1.79) & 3.69 & (3.07 4.32)\\
         \rowcolor{lightgray} \emph{Sub-tree} &  && .75 & (.71 .78)  & .43 & (.26 .6) & .43 & (.38 .47) && 1.98 & (1.82 2.14)  & 2.81 & (1.68 3.94) & 2.97 & (2.8 3.13)\\
        \bottomrule
    \end{tabularx}
   \caption{Cosine Similarity and Earth Mover Distance for all the evaluated metrics with resampling parameter $\epsilon=$0.1 and $\epsilon=$0.2, with trees trimmed to depths 5, 10 and 15. We run the experiments 10 times with random initial vectors and reported the mean values (M) and 95 \% confidence intervals (CI) for VesselVAE trained on synthetic data generated with~\cite{hamarneh2010vascusynth}.We attribute the larger performance drop on Vascusynth, especially at greater depths to a mismatch between VesselVAE’s inductive biases and the synthetic nature of the data. The VesselVAE architecture was designed and tuned for anatomical realism based on real datasets (Aneurisk and Intra), whereas Vascusynth is generated via a parametric algorithm. We hypothesize, the model is being pushed to generalize beyond the structural patterns it was intended to capture.} 
   \label{tab:vascu}
\end{table*}

\end{document}

%% file: pre.tex
\usepackage{dsfont}
\usepackage{amsthm}
\usepackage{overpic}
\usepackage{contour}
\contourlength{1pt}
\usepackage{xspace}
\usepackage{sidecap}
\usepackage{rotating}
\usepackage[normalem]{ulem}
\usepackage{amsmath}
\usepackage{xcolor,colortbl}
\usepackage{tabularx}
\usepackage{soul}
\usepackage{url}
\usepackage{amsfonts}
\usepackage{comment}
\usepackage{booktabs}
\usepackage{etoolbox}
\usepackage{cancel}

\newcommand{\pau}[1]{\textcolor{blue}{#1}}

\newlength{\indentlaenge}
